\documentclass[twocolumn,showpacs,prl,amsmath,amssymb]{revtex4}
\usepackage{graphicx}
\usepackage{dcolumn}
\usepackage{bm}
\usepackage{threeparttable}

\begin{document}

\title{Structure, Stability, Edge States and Aromaticity of Graphene Ribbons}
\author{Tobias Wassmann}
\author{Ari P. Seitsonen}
\author{A. Marco Saitta}
\author{Michele Lazzeri}
\author{Francesco Mauri}
\affiliation{IMPMC, Universit\'{e} Paris 6 et 7, CNRS, IPGP, 140 rue de
Lourmel, 75015 Paris, France}
\date{\today}

\begin{abstract}
We determine the stability, the geometry, the electronic and
magnetic structure of hydrogen-terminated graphene-nanoribbons
edges as a function of the hydrogen content of the environment by
means of density functional theory. Antiferromagnetic zigzag
ribbons are stable only at extremely-low ultra-vacuum pressures.
Under more standard conditions, the most stable structures are the
mono- and di-hydrogenated armchair edges and a zigzag edge
reconstruction with one di- and two mono-hydrogenated sites. At
high hydrogen-concentration ``bulk'' graphene is not stable and
spontaneously breaks to form ribbons, in analogy to the
spontaneous breaking of graphene into small-width nanoribbons
observed experimentally in solution. The stability and the
existence of exotic edge electronic-states and/or magnetism is
rationalized in terms of simple concepts from organic chemistry
(Clar's rule).
\end{abstract}

\maketitle

While two-dimensional
graphene~\cite{journals/Science/Novoselov2004} exhibits
fascinating properties such as a relativistic massless-dispersion
for its charge carriers and ballistic transport on large
distances, its gapless spectrum makes it unsuitable for direct
application as a channel in field-effect transistors and
other semiconducting devices. This problem might be overcome, and
thus open the way to a breakthrough carbon-based electronics, by
designing graphene
nanoribbons (GNRs)~\cite{journals/PhysRevB/Nakada1996a} in which
the lateral quantum confinement opens an electronic gap, function of
the ribbon width. Recent
works~\cite{journals/PhysRevLett/Han2007a,journals/PhysicaE/Chen2007}
have shown that the use of lithographic patterning of graphene
samples can yield GNRs. However, the reported ribbons have large widths,
between 15 and 100 nm, small electronic gap, up to 200 meV,
and are characterized by a significant egde-roughness.
For practical applications, larger
gaps, and correspondingly narrower and smoother-edged GNRs (width
$w\lesssim$~10 nm), are desirable. An alternative experimental
route~\cite{journals/science/Li2008a,wang08}, that makes use of
chemical methods such as solution-dispersion and sonication, has
shown that graphene sheets spontaneously break into ribbons of
narrow width and smooth edges. Ideally, a combination of
the two methods could lead to the design of GNRs having edges of
controlled orientation, spatial localization, and electronic
properties.

In this regard, the knowledge of the structural and energetic
properties of the possible edges, as well as of their
thermodynamic stability, is crucial to achieve the experimental
control necessary for technological applications~\cite{wang08}.
A huge
number of theoretical works has recently appeared in the
literature on GNRs~\cite{pisani07,journals/PhysRevB/Yamashiro2003a,
journals/Nature/Louie2006a,journal/ApplPhysLett/Kan2007a,
journals/PhysRevLett/Yazyev2008a,okada2008,koskinen-2008,Huang2008}.
In those studies, ribbons are usually chosen
with the two fundamental edge geometries, zigzag and
armchair, and dangling bonds are saturated with a single hydrogen
per carbon atom. Zigzag GNRs have caused a
great resonance in the scientific community, due to their
surprising electronic and magnetic properties. In fact, zigzag GNRs
feature magnetism due to spin-polarized electronic states
localized on the edge
~\cite{pisani07,journals/PhysRevB/Yamashiro2003a} and they possibly
turn to half-metal under high external electric
fields~\cite{journals/Nature/Louie2006a,
journal/ApplPhysLett/Kan2007a}. All this has fired the hope to use
GNRs in future
spintronic-devices~\cite{journals/PhysRevLett/Yazyev2008a,
journals/Nature/Louie2006a}, even if the robustness of the spin polarization
in presence of defects has been questioned~\cite{Huang2008}.
Surprisingly, no attempt has been
made so far to study the energetics and thermodynamics of ribbon edges
having different hydrogen terminations, as a function of the chemical
potential typical of the experimental conditions.

In this work we report density functional theory (DFT)
calculations of the energetics and structure of various
hydrogen-terminated graphene-nanoribbons edges as a function of
the hydrogen content of the environment. We show that magnetic
nanoribbons are stable only at extremely low
hydrogen-concentrations, challenging to obtain experimentally.
They are very reactive and thus unlikely to be stable at ambient
conditions. The most stable structures at reasonable
hydrogen-concentrations are non-magnetic and have a very low
chemical reactivity. They are thus likely to be metastable well
beyond their region of thermodynamical stability. We also show
that at high hydrogen-concentrations graphene is not stable and
spontaneously breaks to form hydrogenated edges.  This is
analogous to the spontaneous breaking of graphene into small-width
nanoribbons observed experimentally in organic compounds solution
~\cite{journals/science/Li2008a,wang08}. Finally, the stability
and the existence of exotic electronic edge-states and/or
magnetism are fully rationalized in terms of simple concepts from
organic chemistry (Clar's rule).

\begin{figure*}[t]
\includegraphics[width=175mm]{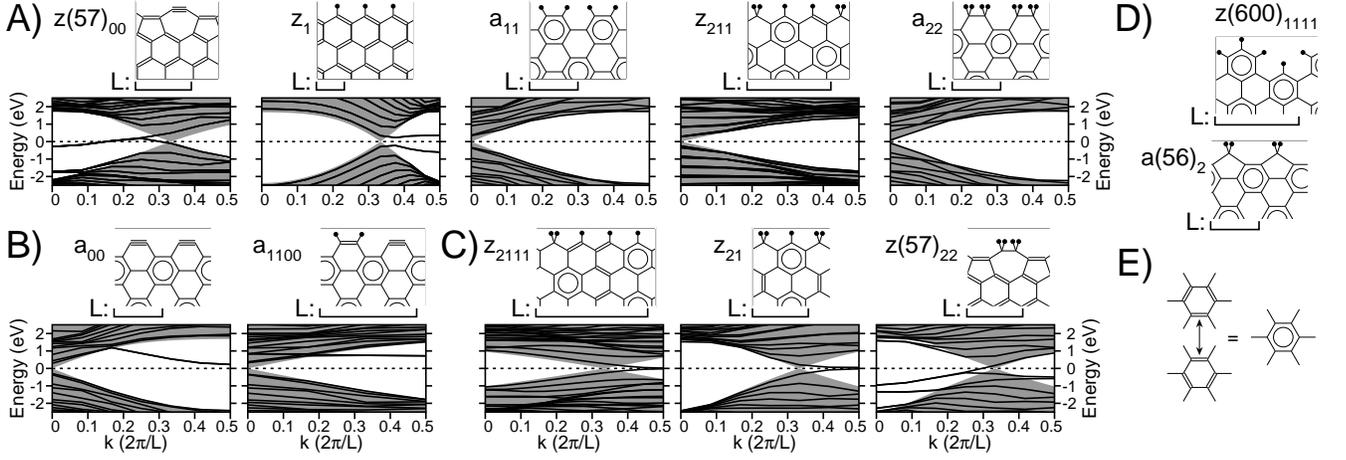}
\caption{
A: Scheme and electronic band-structure of the five most stable
hydrogen-terminated edges of a graphene nanoribbon.
Carbon-carbon bonds are represented with the standard notation, while
hydrogen atoms are the filled circles.
The structures are periodic along the ribbon-edge with periodicity L.
The gray area corresponds to the electronic-bands allowed in
``bulk'' graphene. The dashed line is the Fermi level.
B, C: Other stable armchair and zigzag terminations.
D: Further examples of possible edges.
E: Standard representation of the benzenoid aromatic carbon ring.
}
\label{fig:bands}
\end{figure*}

We studied GNRs with different edge-terminations using periodic super-cell
geometries with lattice-parameter $L$ (periodicity) along the edge,
the two edges having the same configuration.
Details of the total-energy calculations are in note
~\cite{CompDetails}.
Zigzag edges are denoted as z$_{\mathrm{n_1}\mathrm{n_2}...\mathrm{n_X}}$
where $\mathrm{n_i}$ stands for the number of hydrogen atoms on a given
site, and X is the number of adjacent edge-sites within the periodicity $L$.
Armchair edges are denoted as
a$_{\mathrm{m_1}\mathrm{m_2}}$
(a$_{\mathrm{m_1}\mathrm{m_2}\mathrm{m_3}\mathrm{m_4}}$)
for supercells containing one (two) hexagon column(s), where
m$_{\mathrm i}$ indicates the number of hydrogen atoms bonded to
the $i^{th}$ carbon site.
Some examples are in Fig.~\ref{fig:bands}.
Other structures which do not fit this notation,
z(57)$_{ij}$, z(600)$_{ijkl}$, and a(56)$_i$, are also in Fig.~\ref{fig:bands}.
Again, the subscripts indicate the
number of hydrogens on a given site.

In Tab.~\ref{tab:1} we report
the computed zero-temperature edge formation-energy per length:
\begin{equation}
{\cal E}_{H_2} = \frac{1}{2L}\left( E^{ribb} - N_C E^{blk}
-\frac{N_H}{2}E_{H_2}\right),
\label{eq1}
\end{equation}
where $E^{ribb}$, $E^{blk}$ and $E_{H_2}$ are the total energy
of the ribbon super-cell, of one atom in ``bulk'' graphene
and of the isolated H$_2$ molecule~\cite{zero-point}.
$N_C$ ($N_H$) are the number
of carbon (hydrogen) atoms in the super-cell.
${\cal E}_{H_2}$ can be used to determine the stability of different
structures as a function of the experimental conditions~\cite{quian88}.
In presence of molecular H$_2$ gas, at a given chemical
potential $\mu_{H_2}$,
the relative stability is obtained by comparing
$G_{H_2}={\cal E}_{H_2} -\rho_H\mu_{H_2}/2$, where $\rho_H=N_H/(2L)$.
At the absolute temperature $T$ and for a partial H$_2$ pressure $P$~\cite{bailey07},
\begin{equation}
\mu_{H_2} = H^\circ(T) - H^\circ(0) - TS^\circ(T) + k_BT \ln\left(\frac{P}{P^\circ}\right),
\label{eq2}
\end{equation}
where $H^\circ$ ($S^\circ$) is the enthalpy (entropy) at the pressure
$P^\circ = 1$~bar obtained from Ref.~\cite{H_data}.
In presence of monoatomic-hydrogen gas one should use
$G_H={\cal E}_H -\rho_H\mu_H$, where
${\cal E}_H$ = ${\cal E}_{H_2} - \rho_H\times 2.24$~eV, from DFT.

\begin{figure}[ht!]
\includegraphics[width=80mm]{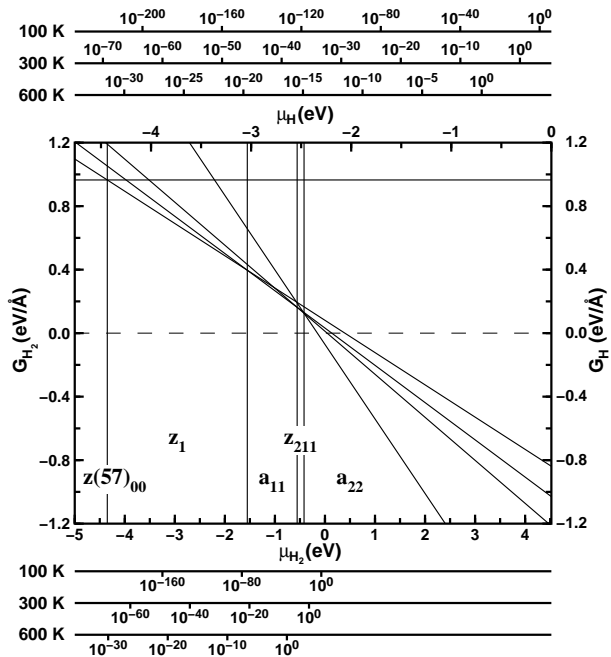}
\caption{Formation energies versus chemical potential for the five most stable
edges. Vertical lines distinguish the stability regions.
The alternative bottom (top) axes show the pressure, in bar, of
molecular H$_2$ (atomic H) gas corresponding to the chemical potentials
at $\textrm{T}=100$, 300, and 600 K.}
\label{fig:edgeenergies}
\end{figure}

\begin{figure}[ht!]
\includegraphics[width=75mm]{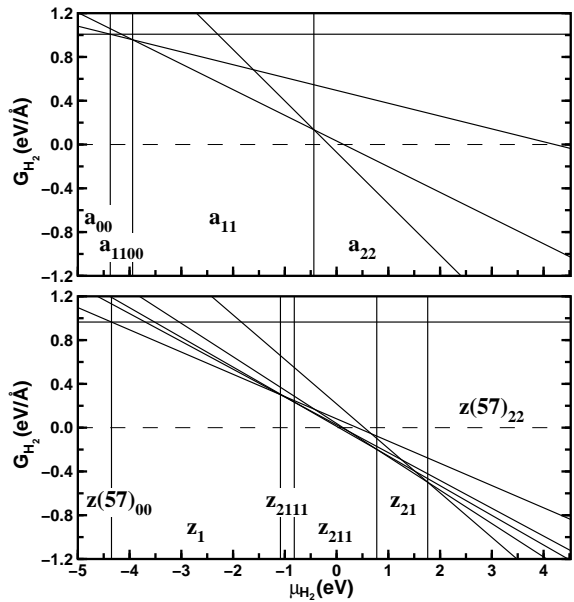}
\caption{Formation energies versus chemical potential for the most
stable  armchair (top) and zigzag (bottom) edges.}
\label{fig:edgeenergies2}
\end{figure}

The most stable structures are thus obtained by comparing
$G_{H_2}$ or $G_H$. $G$ is linear in $\mu$, the slope of the line
being determined by $\rho_H$. For a given value of $\mu$ the
stable structure is the one with the lowest value of $G$, thus, by
increasing $\mu$ ({\it i.e.} going to an environment richer in
hydrogen) the favorable structures will be those with higher
hydrogen-density $\rho_H$. This concept is visualized, as
usual~\cite{quian88}, by plotting $G$ vs. $\mu$ in
Fig.~\ref{fig:edgeenergies}, where we report the five most stable
structures. In Fig.~\ref{fig:edgeenergies2}, we report the same
plot for the most stable edges in the two distinct families of
zigzag and armchair GNRs.

\begin{table}
\caption{Formation energy (Eq.~\ref{eq1}) and
hydrogen density ($\rho_H=N_H/(2L)$) for all the studied edges.}
\begin{tabular*}{\columnwidth}{lcclcc}
&$\rho_H$(\AA$^{-1}$)&${\cal E}_{H_2}$(eV/\AA)&&$\rho_H$(\AA$^{-1}$)&${\cal E}_{H_2}$(eV/\AA) \\
\hline
z(57)$_{00}$    $^\dagger$ & 0.000 & 0.9650 & a(56)$_{0}$ $^*$       & 0.000 &    1.4723 \\
z$_{0}$         $^*$       & 0.000 & 1.1452 & a$_{00}$    $^\circ$   & 0.000 &    1.0078 \\
z$_{100}$       $^*$       & 0.136 & 0.7854 & a(56)$_{1}$ $^\dagger$ & 0.235 &    0.7030 \\
z$_{200}$       $^*$       & 0.271 & 0.7260 & a$_{1100}$  $^\circ$   & 0.235 &    0.4946 \\
z$_{110}$       $^*$       & 0.271 & 0.4306 & a$_{10}$    $^*$       & 0.235 &    0.6273 \\
z(57)$_{11}$    $^\dagger$ & 0.407 & 0.3337 & a$_{11}$    $^\circ$   & 0.469 &    0.0321 \\
z$_{1}$         $^*$       & 0.407 & 0.0809 & a(56)$_{2}$ $^\circ$   & 0.469 &    0.4114 \\
z$_{211111}$    $^*$       & 0.474 & 0.0463 & a$_{21}$    $^*$       & 0.704 &    0.2092 \\
z$_{21111}$     $^*$       & 0.488 & 0.0397 & a$_{2211}$  $^\circ$   & 0.704 & $-$0.0163 \\
z$_{2111}$      $^\dagger$ & 0.508 & 0.0257 & a$_{22}$    $^\circ$   & 0.939 & $-$0.0710 \\
z$_{211}$       $^\circ$   & 0.542 & 0.0119 & & & \\
z(600)$_{1111}$ $^\circ$   & 0.542 & 0.0459 & & & \\
z$_{21}$        $^\dagger$ & 0.610 & 0.0382 & & & \\
z$_{221}$       $^*$       & 0.678 & 0.1007 & & & \\
z$_{2}$         $^*$       & 0.813 & 0.2224 & & & \\
z(57)$_{22}$    $^\dagger$ & 0.813 & 0.2171 & & & \\
\end{tabular*}
\begin{tabular*}{\columnwidth}{rrrr}
$\dagger$ & Non magnetic,   &    metallic edges, & non aromatic \\
$*$       & Magnetic,       &    metallic edges, & non aromatic \\
$\circ$   & Non magnetic,   & non-metallic edges, & aromatic \\
\end{tabular*}
\label{tab:1}
\end{table}

Some important conclusions can be driven from Fig.~\ref{fig:edgeenergies}.
First, the magnetic z$_1$ edge, widely studied
as a potential component for future spintronic devices
~\cite{journals/PhysRevLett/Yazyev2008a,journals/Nature/Louie2006a},
and the z(57)$_{00}$, proposed in~\cite{koskinen-2008},
are stable only at extremely low hydrogen-concentrations.
Second, at reasonable hydrogen pressure (at ambient conditions, the
partial H$_2$ pressure in air is $\sim 5\times 10^{-7}$~bar) the most
stable edges are a$_{11}$, z$_{211}$, and a$_{22}$.
Finally, Fig.~\ref{fig:edgeenergies} shows that graphene is not stable
at high hydrogen-concentration ($G$ becomes
negative) and spontaneously breaks to form hydrogenated edges.
Similarly, it has been experimentally observed that
small-width nanoribbons spontaneously form in organic compounds
solutions ~\cite{journals/science/Li2008a,wang08}.

We indicate in Tab.\ \ref{tab:1}, along with the energetics,
whether the proposed edges are magnetic or not, \emph{i.e.}
whether a finite magnetization density is observed within DFT calculations.
In Fig.~\ref{fig:edgeenergies}, we also report the electronic bands
of the ribbons and we show the bands of two-dimensional ``bulk''
graphene projected on the one-dimensional Brillouin zone of
the edge (gray area in Fig.~\ref{fig:edgeenergies}).
As currently done in surface physics, we call
``edge states'' (ESs) those electronic states localized
at edges, having an energy which is not allowed in the bulk ({\it i.e.}
those bands which are outside the gray area in Fig.~\ref{fig:edgeenergies}).
Such states are localized near the edge.
Moreover, in Tab.~\ref{tab:1} we indicate as ``metallic edges'', those having
partially-filled ESs, \emph{i.e.} having ESs which, in absence of
magnetic order, cross the Fermi level, and are thus only partially
occupied. For a system with partially filled ESs, in a ribbon,
it might be energetically
convenient to split those bands by inducing a magnetic
order between the two ribbon-edges and thus opening an energy gap,
as in the well-known case of the z$_1$ edge.
Finally, we remind that the chemical reactivity of a given
system can be quantitatively expressed as proportional to the
electronic density of states near the Fermi level.
As a consequence, all the structures having
edge-states crossing the Fermi energy or in its immediate vicinity
(independently on whether they are magnetic or not) are
expected to be extremely reactive, since the edge-states are
available to form chemical bonds.

With these concepts in mind, we remark that the most stable structures
at standard and high hydrogen-concentrations, namely a$_{11}$, z$_{211}$,
and a$_{22}$, have no edge states, in agreement with the experimental
observation of large semiconducting gaps in narrow ribbons~\cite{wang08}.
This implies that,
besides being non-magnetic, they have a very low chemical
reactivity. As a consequence, once formed, those edges are likely
to be metastable well beyond their respective windows of thermodynamical
stability in Fig.\ \ref{fig:edgeenergies}.
On the other hand, magnetic edges, as the well studied z$_1$, are
stable only at extremely low hydrogen-concentrations and are also very
reactive and thus unlikely to be stable at ambient conditions.

We now show that the electronic structure of the edges can be
interpreted in terms of aromaticity and according
to Clar's rule, well known in organic chemistry since the
60's~\cite{books/Clar1964a, books/Clar1972a}.
The fundamental stability-criterion for
hydrocarbons is that each carbon atom must have four saturated bonds.
In organic chemistry, the so-called benzenoid aromatic ring
can be defined as a resonance
(due to the delocalization of $\pi$ electrons over the
ring~\cite{journals/ChemRev/Wu2007, journals/ChemRev/Watson2001a})
between 2 hexagonal rings with alternating single and double bonds as in
Fig.\ \ref{fig:edgeenergies}-E.
For such a structure, all the bonds sticking out of the hexagon are
single bonds and, as a consequence, two benzenoid rings cannot
be adjacent.
The energetic stability of a structure increases with the number
of possible resonances. As a consequence, since a structure
containing $n$ benzenoid rings displays $2^n$ possible resonances,
the most stable isomer of a given hydrocarbon is the one that maximizes the
number of benzenoid rings (Clar's rule
~\cite{books/Clar1964a, books/Clar1972a}).
Bulk graphene is the ideal case of aromaticity since there are three
equivalent representations in which all the $\pi$ electrons belong to
a benzenoid ring and no double-bonds are present.
In such a representation 1/3 of the graphene carbon-hexagons are
benzenoid rings.

The edge structures of Fig.\ \ref{fig:bands} are displayed
in a representation that maximizes the number of benzenoid rings.
However, the resulting ring-density for some of the structures
(e.g. the z$_{21}$ or z$_{2111}$) is lower
than the ideal graphene 1/3 density.
We analyzed all the edges of Tab.\ \ref{tab:1} and we indicate
as ``aromatic'' those structures that have a density
of benzenoid rings of 1/3 (e.g. the a$_{11}$, z$_{211}$ and a$_{22}$).
We note that {\it all the aromatic structures do not have
partially-filled edge-states (are non-metallic) and vice-versa}.
This empirical rule is valid for all the structures presently studied
and can be understood in terms of the Clar's rule.
In fact, whenever the ring-density near the edge is smaller than 1/3,
there is a competition between the bulk,
where aromaticity prevails with an exponential ($2^n$) weight, and
the edge. It is precisely this competition that, imposing the
ring density of the bulk, forces some edge carbon-atoms to have
less or more than 4 saturated bonds and, thus, originates the electronic
``defects'', {\it i.e.} the edge states.
The z$_{211}$ and the z(600)$_{1111}$
ribbons are the only zigzag edges compatible with the aromaticity of the
bulk. These edge configurations have in fact the same 1/3 periodicity
along the zigzag edge as the benzenoid rings of ``bulk'' graphene.
Moreover, mono- and di-hydrogenated armchair
edges are both compatible with the 1/3 aromaticity of graphene, whereas
asymmetrically passivated edges, such as a$_{21}$, are not.
Notice that, {\it for a given H density, the most stable edges are aromatic,
if an aromatic structure is compatible with such a concentration}.
The stability of fully-aromatic edges is confirmed by the fact
that polycyclic aromatic hydrocarbons whose structure can be fully
represented by benzenoid rings (without double bonds) show
particularly high stability, high melting points, low
reactivity~\cite{journals/ChemRev/Wu2007}, and have, whenever
possible, a$_{11}$ or z(600)$_{1111}$
edges~\cite{journals/ChemRev/Watson2001a,journals/ChemA/Muller1998}.

Concluding, the knowledge of the structure and stability of the
possible edges is a crucial issue to control the experimental
conditions of the formation of graphene nanoribbons of desired
properties. Here, we determined the structure and stability of
hydrogen-terminated nanoribbons. For reasonable hydrogen
concentrations the most stable structures are the a$_{11}$,
a$_{22}$ and z$_{211}$, which are not magnetic and do not present
valence electronic states localized at the edge. These structures
are, thus, expected to be non-reactive and metastable well beyond
the region of their thermodynamical stability. In particular, at
high hydrogen concentrations graphene spontaneously breaks into
a$_{22}$ nanoribbons. Our results are rationalized by means of
simple concepts of organic chemistry that can be used to guide the
search for the most stable edges in different chemical
environments.

We thank Th. Greber for discussions.
Calculations were done at IDRIS (project n$^\circ$ 081202).


\begin{thebibliography}{22}
\bibitem{zero-point}
We checked that the inclusion of zero point motion (not considered here
for sake of clarity) does not affect the relative stability of different
edge-structures.
\end{thebibliography}
\end{document}